\documentclass[aps,pra,reprint,superscriptaddress]{revtex4-2}
\usepackage{amsmath}
\usepackage{amssymb}
\usepackage{graphicx}
\usepackage{dcolumn}
\usepackage{bm}
\usepackage{color}
\usepackage{subfigure}
\begin{document}

\title{Quantum-enhanced rubidium atomic magnetometer 
based on Faraday rotation via 795-nm Stokes operator squeezed light
}

\author{Lele BAI $^{1, \# }$, Xin WEN $^{1, 4, \#}$, Yulin YANG $^1$, Lulu ZHANG $^1$, 
Jun HE$^{1, 2, 5}$, Yanhua WANG$^{1, 3, 5}$, and Junmin WANG$^{1, 2, 5, \dagger}$
 \\ \it {$^1$ State Key Laboratory of Quantum Optics and Quantum Optics Devices, and Institute of Opto-Electronics, Shanxi University, Tai Yuan 030006, People’s Republic of China}\\
\it {$^2$ Department of Physics, School of Physics and Electronic Engineering, Shanxi University, Tai Yuan 030006, People’s Republic of China}\\
\it {$^3$ Department of Opto-Electronic Engineering, School of Physics and Electronic Engineering, Shanxi University, Tai Yuan 030006, People’s Republic of China}\\
\it {$^4$Department of Physics, Tsinghua University, Beiing 100084, People’s Republic of China}\\
\it {$^5$Collaborative Innovation Center of Extreme Optics, Shanxi University, Tai Yuan 030006, People’s Republic of China}}
\email{$\# $ These authors contributed equally to this work. \\$\dagger$ Corresponding author, E-mail: wwjjmm@sxu.edu.cn}

\date{\today}

\begin{abstract}
With the help of Stokes operator $\widehat{S}_2$ squeezed state (also called polarization squeezed state (PSS)) of 795-nm light, rubidium-87 ($^{87}$Rb) atomic magnetometer based on Faraday rotation has been implemented and characterized. The PSS of Stokes operator $\widehat{S}_2$ of 795-nm light has been prepared by means of coherently combining the polarization coherent state (PCS) of a linearly p-polarized bright 795-nm light beam and a linearly s-polarized squeezed vacuum state (SVS) generated by a 397.5-nm ultraviolet laser pumped sub-threshold optical parametric oscillator (OPO) with a PPKTP bulk crystal inside the OPO cavity. PSS with a squeezing level of -3.7$\pm$0.3 dB has been achieved around the analysis frequency of 10 kHz. At different transitions of D$_1$ line, various frequency detuning, and reasonable atomic vapor cell’s temperature, Faraday rotation has been measured and compared. To decrease absorption (scattering) losses and the back-action from atomic spin noise to the probe beam’s polarization noise for maintaining the quantum properties of PSS of Stokes operator $\widehat{S}_2$ of 795-nm light, we had to run our magnetometer with $^{87}$Rb vapor cell’s temperature below 60 ℃, at which the PSS was almost destroyed. The sensitivities of magnetic field measurement were characterized via measuring signal-to-noise ratio of the alternating current (AC) calibrated magnetic field signal with a balanced polarimeter. Under the conditions of the atomic number density of $\sim$5.8×10$^{10}$ /cm$^3$ (T$\sim$ 40 ℃) and the probe beam with a detuning of $\sim$ - 400 MHz relative to the 5S$_{1/2}$ (F$_g$=2) - 5P$_{1/2}$ (F$_e$=1) transition of $^{87}$Rb D$_1$ line, a typical sensitivity of $\sim$19.5 pT/Hz1/2 has been achieved employing PSS of Stokes operator $\widehat{S}_2$ as the probe, compared with a sensitivity of $\sim$28.3 pT/Hz$^{1/2}$ using PCS as the probe. We preliminarily demonstrated that the quantum-enhanced sensitivity in a Faraday-rotation-based $^{87}$Rb atomic magnetometer with the help of PSS of 795-nm light.

\end{abstract}

\maketitle
\section{Introduction}
The squeezed state of light, one kind of quantum optical resources, has a reduced noise level in a certain orthogonal component than the corresponding shot noise level (SNL). It can be widely used in many precision measurement fields to extract weak signal, such as the gravitational wave detection [1, 2], magnetometers [3-7], and spin noise spectroscopy [8], which is attributed to the fact that the SNL can be broken through and the signal-to-noise ratio (SNR) can be improved when the system reaches the limit of classical measurement. According to the principle of quantum optics, Stokes operator squeezed state of light, also called polarization squeezed state (PSS) of light, can be employed to improve the sensitivity of magnetic field measurement. In 2010, Wolfgramm et al. [3] employed PSS of light as the probe of a rubidium 87 ($^{87}$Rb) atomic magnetometer based on Faraday rotation (FR) and demonstrated the sensitivity improvement from $\sim$ 46 nT/Hz$^{1/2}$  to $\sim$ 32 nT/Hz$^{1/2}$  around the analysis frequency of 120 kHz with a squeezing level of -3.2 dB, which is attributed to the far detuned probe laser relative to the $^{87}$Rb transition, so the quantum properties of PSS can be well maintained. In 2012, Horrom et al. [4] employed PSS of 795 nm light prepared by using of the polarization self rotation (PSR) in a $^{87}$Rb vapor cell, whose squeezing level is strongly dependent on frequency detuning, to demonstrate the sensitivities at analysis frequency of 500 kHz under different atomic number densities, but it is difficult to highlight its quantum enhancement properties because of the limited squeezing level of only 2 dB, especially with a high atomic number density or the other frequency detuning. With a typical atomic number density of $\sim$2.1×10$^{10}$/cm$^3$, the sensitivity has been improved from $\sim$ 80 pT/Hz$^{1/2}$ to $\sim$ 65 pT/Hz$^{1/2}$. Novilova et al. [5] demonstrated a different type of fundamental noise in atomic sensors by using PSS with the squeezing level of about -2 dB and a atomic vapor cell with paraffin wall coating, which can reduce the excess optical quantum noise by increasing the purity of the squeezed state. They also proved that even though the measured magnetic-field sensitivity of their device continues to improve with the growth of the atomic density, but it improves at a lower rate than estimated only by the shot-noise-limited sensitivity. In 2014, Otterstrom et al. [6] presented a magnetometer that produces two mode-squeezed states from a four-wave mixing pr℃ess while simultaneously performing in situ sub-shot noise magnetometry in a single vapor cell, which was very different from the way we studied, and the magnetic field is measured by using a two-photon degenerate single beam of PSS passing through a pure $^{87}$Rb atomic vapor cell based on the Faraday rotation mechanism. Typically, the sensitivity can be improved from $\sim$33.2 pT/Hz$^{1/2}$ to $\sim$19.3 pT/Hz$^{1/2}$ at the analysis frequency of $\sim$700 kHz by the intensity difference squeezed state of light with squeezing level of -4.7dB. In 2018, Li et al. [7] enhanced the sensitivity at MHz analysis frequency by using 1064 nm phase squeezed light as probe in a magnetometer based micro-cavity. Typically, when the analysis frequency is $\sim$ 8.6 MHz, the sensitivity can be improved from $\sim$ 35.9 nT/Hz$^{1/2}$ to $\sim$ 29.2 nT/Hz$^{1/2}$. The above quantum enhancement measurements of magnetic fields were all carried out at high analysis frequency where the squeezing is easier to detect, however, we chose the analysis frequency of 10 kHz for characterization, which requires us to overcome the difficulty to prepare PSS with high squeezing level at the audio frequency. Meanwhile, for magnetometer based on FR mechanism, its sensitivity will get worse due to the high background noise at such low frequency. To be fair, various systems with different experimental parameters, including magnetic shielding device, laser frequency, atomic number density, type of atomic vapor cell and the analysis frequency of calibrating magnetic field, their respective sensitivities probed by polarization coherent state (PSS) of light are generally different. In addition, the way to prepare PSS and their corresponding squeezing levels at different analysis frequencies are also diverse, so the quantum-enhanced effect under the action of PSS will be not the same. In addition, in order to obtain higher sensitivity, the preparation of atomic spin squeezing is also necessary [9,10]. There are many other schemes that have been implemented for atomic magnetometers. Among them, the atomic magnetometer based on spin-exchange relaxation-free state has been optimized with the best state-of-the-art sensitivity of $\sim$ 0.16 fT/Hz$^{1/2}$ [11], while the atomic magnetometer using the coherent population trapping effect has been optimized with a sensitivity of $\sim$ 12 pT/Hz$^{1/2}$ [12].

In this paper, we present an atomic magnetometer based on the FR mechanism, only single laser beam called PCS or PSS of light is used for both pumping and probing. Here, squeezed vacuum state (SVS) with squeezing level of -4.0$\pm$0.3 dB at the analysis frequency of 10 kHz is obtained via the optical parametric oscillator (OPO). Quantum noise l℃king (QNL) technology [13] is used to l℃k the squeezed phase in real time, and Stokes operator $\widehat{S}_2$ PSS with squeezing level of -3.7$\pm$0.3 dB can be eventually prepared. The responses of the probe laser that resonates with different atomic transition lines to the external magnetic field are measured. The constraints of atomic number density and laser frequency detuning on the quantum properties of PSS after passing through the atomic ensemble and their respective influences on the sensitivity of magnetic field measurement are analyzed, which proves that the quantum-enhanced effect of PSS can only be demonstrated under certain conditions. However, the corresponding experimental parameters are often not the best choice for PCS. Here, we limit the atomic number density in order to preserve the quantum properties of PSS and initially demonstrate its benefits for measuring the sensitivity of magnetic field. Typically, given an atomic number density of $\sim$ 5.8×10$^{10}$/cm$^3$ (T$\sim$ 40 ℃) and laser frequency detuning of -400 MHz relative to the 5S$_{1/2}$ (Fg=2) - 5P$_{1/2}$ (F$_e$=1) hyperfine transition of D$_1$ line of $^{87}$Rb, the sensitivity of the AC magnetic field at 10 kHz has been improved from $\sim$28.3 pT/Hz$^{1/2}$ to $\sim$19.5 pT/Hz$^{1/2}$ with the help of PSS.

\section{Theoretical analysis}
The mechanism of Faraday rotation can be interpreted as a consequence of the light-induced dichroism in an atomic ensemble when a longitudinal magnetic field is applied parallel to the direction of propagation of probe laser, which can be described in three stages: optical pumping with resonant or nearly resonant light, atomic spin procession under magnetic filed and optical probing by polarimeter. An atomic ensemble is polarized by shining a linearly polarized light on it and shows linear dichroism. When there is a longitudinal magnetic field, the magnetic quadrupole moment of the atoms will process around the magnetic field, deflecting the linear dichroism axis of the ensemble. At this point, its polarization direction is no longer parallel to the polarization of the laser beam and the vertical and horizontal components of linearly polarized light are absorbed to different degrees, resulting in the rotation of the polarization plane. The form of rotation angle depending on magnetic field is a dispersion-like curve and its magnitude is proportional to the magnetic field near-zero. We built a model to depict the quantum noise distribution of PSS and PCS on the Poincare sphere and the trajectory of their respective polarization plane in the FR mechanism, as shown in figure 1. Assuming that the laser fields at the initial position is horizontally polarized, it can be converted into linearly polarized light 45° from the horizontal as it rotates 90° around the equator.
\begin{figure}
\centering
\subfigure[]{\label{fig:subfig:a}
\includegraphics[scale=0.7]{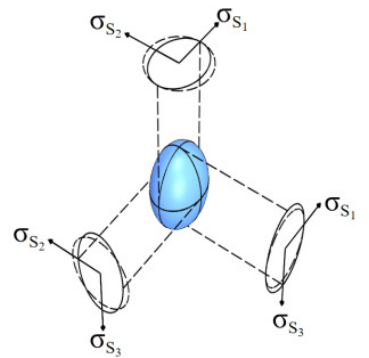}}
\hspace{0.01\linewidth}
\subfigure[]{\label{fig:subfig:b}
\includegraphics[scale=0.7]{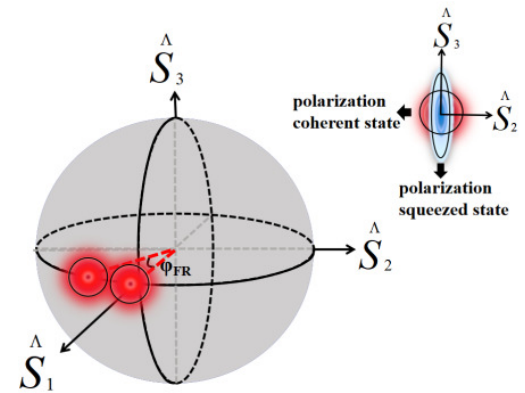}}
\vfill
\subfigure[]{\label{fig:subfig:a}
\includegraphics[scale=0.7]{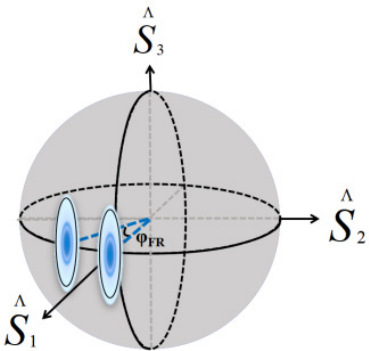}}
\hspace{0.01\linewidth}
   \caption{(a) Quantum polarization noise ellipsoid of PSS of light and its projection along three vertical directions. Quantum noise of Stokes operator $\widehat{S}_2$ is squeezed, while Stokes operator $\widehat{S}_3$ is anti-squeezed. The surface of the blue ellipsoids defines the standard deviation of the noise normalized to the SNL (${{\sigma }_{{{S}_{1}}}}\text{=}\sqrt{{{V}_{\text{1}}}}{{\sigma }_{{{S}_{2}}}}\text{=}\sqrt{{{V}_{\text{2}}}}{{\sigma }_{{{S}_{3}}}}\text{=}\sqrt{{{V}_{\text{3}}}}$), where $V_1, V_2$, and $V_3$ represent the noise variance of each orthogonal component, respectively. (b) Polarization rotation is depicted on a Poincaré sphere based on FR-type atomic magnetometer with PCS of light. The red sphere and the blue ellipse on the upper right indicates the noise shapes of PCS and PSS in the same rectangular coordinate system, respectively. (c) In the case of PSS as the case of (a), the recognizable Faraday rotation angle of $\varphi_{FR}$ is much smaller because the quantum noise of Stokes operator $\widehat{S}_2$ is squeezed, therefore it is more sensitive to external magnetic field.}
\end{figure}

The initial polarization of PCS and PSS are both along $\widehat{S}_1$, and the balanced polarimeter measures the value of $\widehat{S}_2$ from the intensity [14, 15],
\begin{equation}
{{\hat{S}}_{2}}={{\hat{I}}_{+45}}-{{\hat{I}}_{-45}},
\end{equation}
where ${{\hat{I}}_{+45}}$ and ${{\hat{I}}_{-45}}$ represent the laser intensity operators of the two arms of the balanced detector.

The polarization rotation is expressed as [4],

\begin{equation}
\hat{S}_{2}^{\left( out \right)}=\hat{S}_{2}^{\left( in \right)}+{{\hat{S}}_{1}}{{\varphi }_{FR}},
\end{equation}
where ${{\hat{S}}_{1}}={{\hat{I}}_{H}}-{{\hat{I}}_{V}}$ is the difference of laser intensity between the horizontal polarization and vertical polarization, and $\hat{S}_{2}^{\left( in \right)}$and $\hat{S}_{2}^{\left( out \right)}$are input and output states of operator $\hat{S}{}_{2}$ respectively, ${{\varphi }_{FR}}$ is the corresponding rotation angle of polarization.  

The noise variance of the output measured by spectrum analyzer is
\begin{equation}
\operatorname{var}[\hat{S}_{2}^{\left( out \right)}]=\operatorname{var}[\hat{S}_{2}^{\left( in \right)}]+\operatorname{var}[{{\hat{S}}_{1}}{{\varphi }_{FR}}],
\end{equation}
where the first and second terms on the right-hand side are the shot noise of photons and the atomic spin noise respectively, independent of each other. When \[\operatorname{var}[\hat{S}_{2}^{\left( out \right)}]=\operatorname{var}[{{\hat{S}}_{1}}{{\varphi }_{FR}}]\], it means that the input laser field is PCS exhibiting Poissonian statistics, and when \[\operatorname{var}[\hat{S}_{2}^{\left( out \right)}]<\operatorname{var}[{{\hat{S}}_{1}}{{\varphi }_{FR}}]\], it means PSS exhibiting sub-Poissonian statistics. Therefore, it is very reasonable and promising that the background noise can be overcome by using PSS, realizing the quantum enhancement of sensitivity.

\section{Experimental setup}
A continuous-wave narrow-linewidth Ti: Sapphire laser (M Squared, Model SolsTis) tuned to 795 nm rubidium D1 line is used as a laser source. The second-harmonic generation cavity with an LBO crystal inside and OPO cavity with a PPKTP crystal inside are combined to produce the s-polarized SVS, a mode cleaning cavity is used to prepare the P-polarized LO oscillator beam (also be called PCS), see figure 2. PSS of light is obtained by synthesizing two orthogonal optical fields with a polarization beam splitter (PBS) cube, which can also be generated in other ways, such as PSR [16] and four-wave mixing [17] in alkali atomic vapor. Differential detection is carried out by a polarimeter consisting of a zero-order 795-nm half-wave plate, PBS with a high extinction ratio ($\sim$ 3300:1) and a balanced detector, at which the noise of the Stokes operator $\hat{S}{}_{2}$ is measured [18, 19]. A 75-mm-long $^{87}$Rb enriched cylindrical atomic vapor cell with 795-nm anti-reflection coating on two windows’ outer surfaces and with no buffer gas, the coils that produce the axial magnetic fields are placed inside a four-layer permalloy ($\mu$-metal) magnetic shielding tank, which can screen the environmental magnetic fields and ensure that the residual magnetic fields at the location of the atomic vapor cell is less than $\sim$ 0.2 nT. 

For both PCS and PSS, the diameter of the beam should be enlarged by the beam expanding telescope (BET), the Gaussian diameter of incident laser beam is $\sim$ 3 mm defined by the laser intensity dropping to the maximum of 1/e$^2$, and corresponding power is $\sim$ 1 mW, which facilitates the full contact between laser and atoms, thus improving the sensitivity of the magnetic field measurement. The noise of the laser field and the sensitivity of AC magnetic field to be measured can be characterized by a radio-frequency spectrum analyzer (RF-SA) and a fast-Fourier-transform dynamic signal analyzer (FFT-DSA), respectively.

In the schemes of calibrating magnetic filed measurement sensitivity, the method of calibration with a known AC magnetic field has high sensitivity, especially for FR scheme with a single probe beam[20, 21]. A calibrated AC magnetic field applied along the propagation direction of the probe laser is generated by a tightly wound solenoid, which is driven by a low-noise current source (Keysight, Model B2961A) with an accuracy of 100 nV/10 fA, the voltage noise output by the polarimeter can be converted to magnetic noise. (Here, the size of AC magnetic field can be measured in magnetic shielding cylinder by using a Hall type magnetometer with characteristics of high accuracy, for weak magnetic fields, it’s value can be calculated by the way of linear extrapolation.) The $SNR$ of AC magnetic field depends on the magnitude of the signal and the background noise of the laser field. The sensitivity expressed by the power spectral density is defined by the ratio between magnitude of the AC magnetic field and the corresponding SNR [22, 23]:
\begin{equation}
\delta B=\frac{\Delta B}{SNR},
\end{equation}

\begin{figure}
\centering
\includegraphics[scale=0.3]{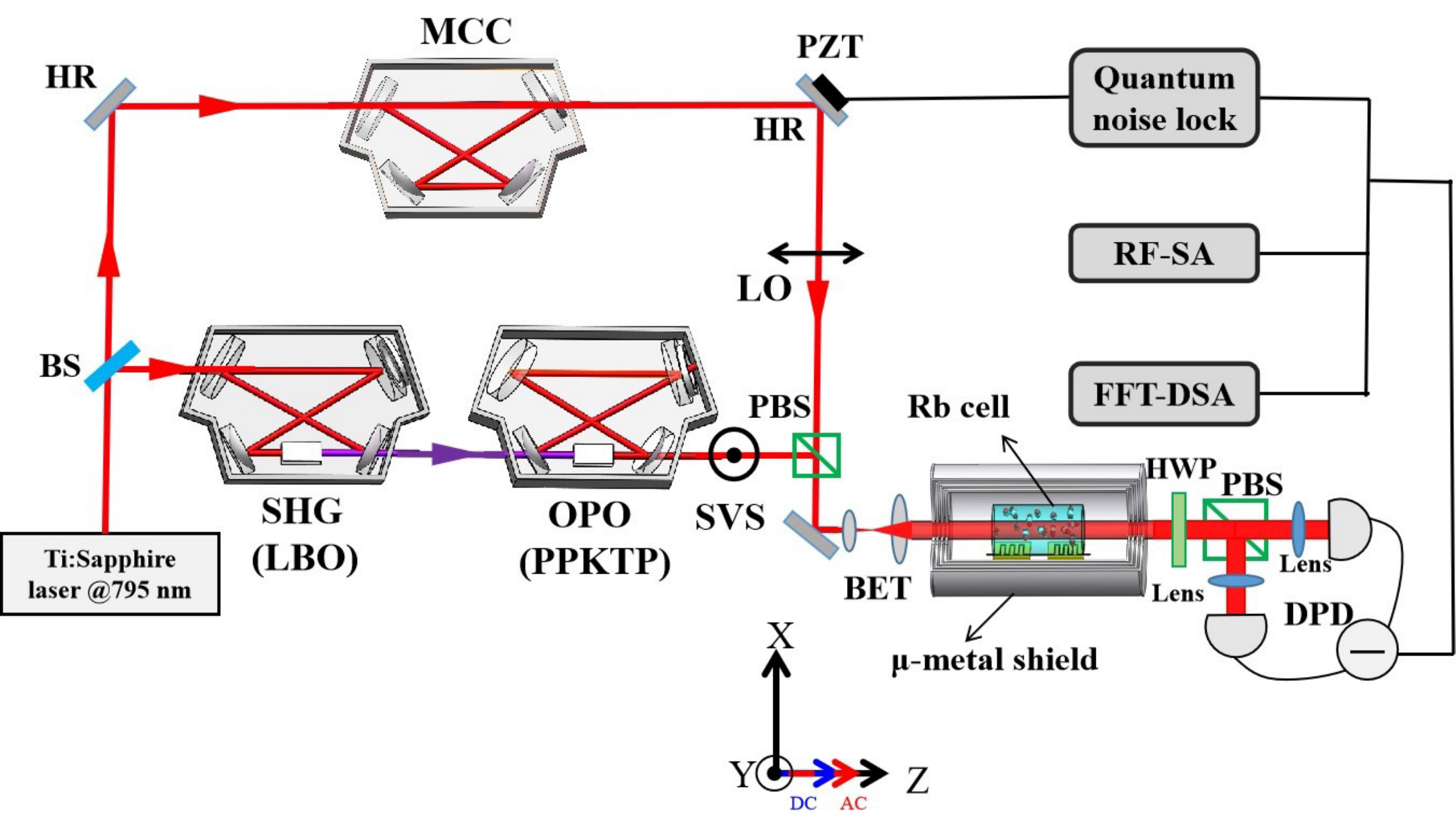}
\caption{(a) Relative noise power spectra of 795-nm SVS of light at the analysis frequency of 10 kHz versus time. The red curve represents the relative noise power normalized to the shot noise level (black line) when the LO’s phase is scanned. The RF spectrum analyzer is set to zero-span mode; the resolution bandwidth (RBW) is set to 1 kHz, while the video bandwidth (VBW) is set to 30 Hz; average is 60 times. (b) Relative noise power spectra of 795-nm PSS state of light at the analysis frequency of 10 kHz versus time. Stokes operator $\widehat{S}_2$ with -3.7$\pm$0.3 dB squeezing is realized when the LO is locked to the squeezing phase via the quantum noise locking scheme. The deterioration of the squeezing level after l℃king is mainly due to the phase noise. (c) Squeezing level of 795-nm polarization squeezed light passing through the $^{87}$Rb-enriched atomic vapor cell at different temperatures versus the laser frequency detuning relative to the 5S$_{1/2}$ (F$_g$ = 2) - 5P$_{1/2}$ (F$_e$ = 1) transition of D$_1$ line of $^{87}$Rb. The green dash line indicates the shot noise level (0 dB). The red circles are for the case without the atomic vapor cell, and the polarization squeezing level is $\sim$ -3.7$\pm$0.3 dB (red solid line). The blue squares are for the case of T$\sim$ 313 K (T$\sim$ 40 ℃) , the pink triangles for the case of T $\sim$ 323 K (T$\sim$ 50 ℃) , and the black stars for the case of T $\sim$ 333 K (T$\sim$ 60 ℃), and all these three sets of data are results for the polarization squeezed light passing through the atomic vapor cell.}
\end{figure}

\section{Preparation of 795-nm Stokes operator $\widehat{S}_2$ polarization squeezed light}
At the analysis frequency of 2 MHz, we have prepared 795-nm SVS with squeezing level of -5.6 dB [18]. However, the preparation of PSS at the audio-frequency is relatively difficult due to the challenging control of system noise and the serious coupling of ambient noise [19, 24], we have demonstrated that the squeezing level of PSS can be improved experimentally by using the laser intensity noise suppression systems based on acousto-optic modulator [25]. In this work, SVS with squeezing level of -4.0$\pm$0.3 dB (the blue dash line) and anti-squeezing level of +7.0$\pm$0.3 dB (the pink dash line) can be obtained at the analysis frequency of 10 kHz by scanning the LO beam’s phase, as shown by the red curve in figure 3 (a); PSS with relative noise power in $\widehat{S}_2$ is $\sim$ -3.7$\pm$0.3 dB (the noise level is about 0.43 times of SNL), as shown by the red line in figure 3 (b).
\begin{figure}
\centering
\subfigure[]{\label{fig:subfig:a}
\includegraphics[scale=0.5]{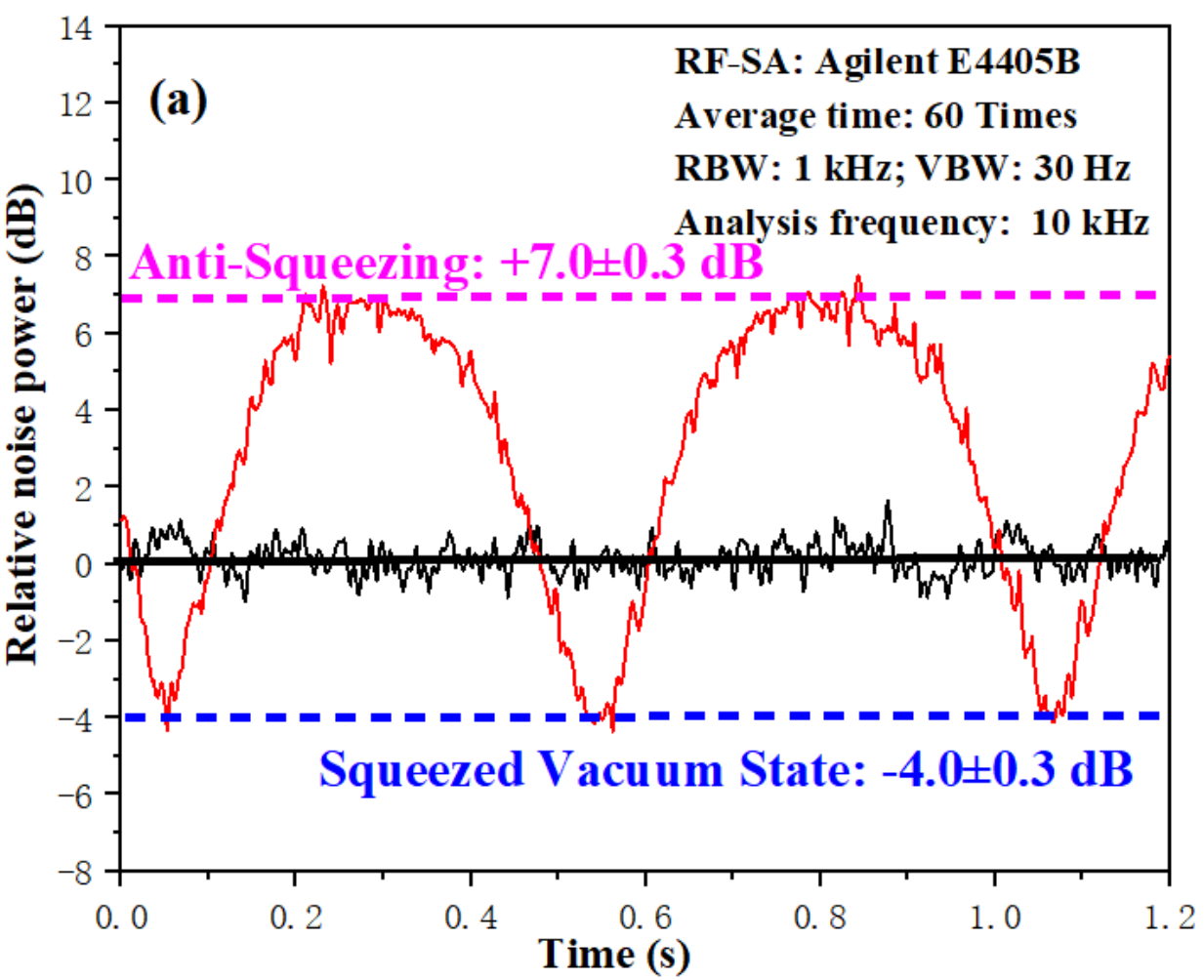}}
\hspace{0.01\linewidth}
\subfigure[]{\label{fig:subfig:b}
\includegraphics[scale=0.5]{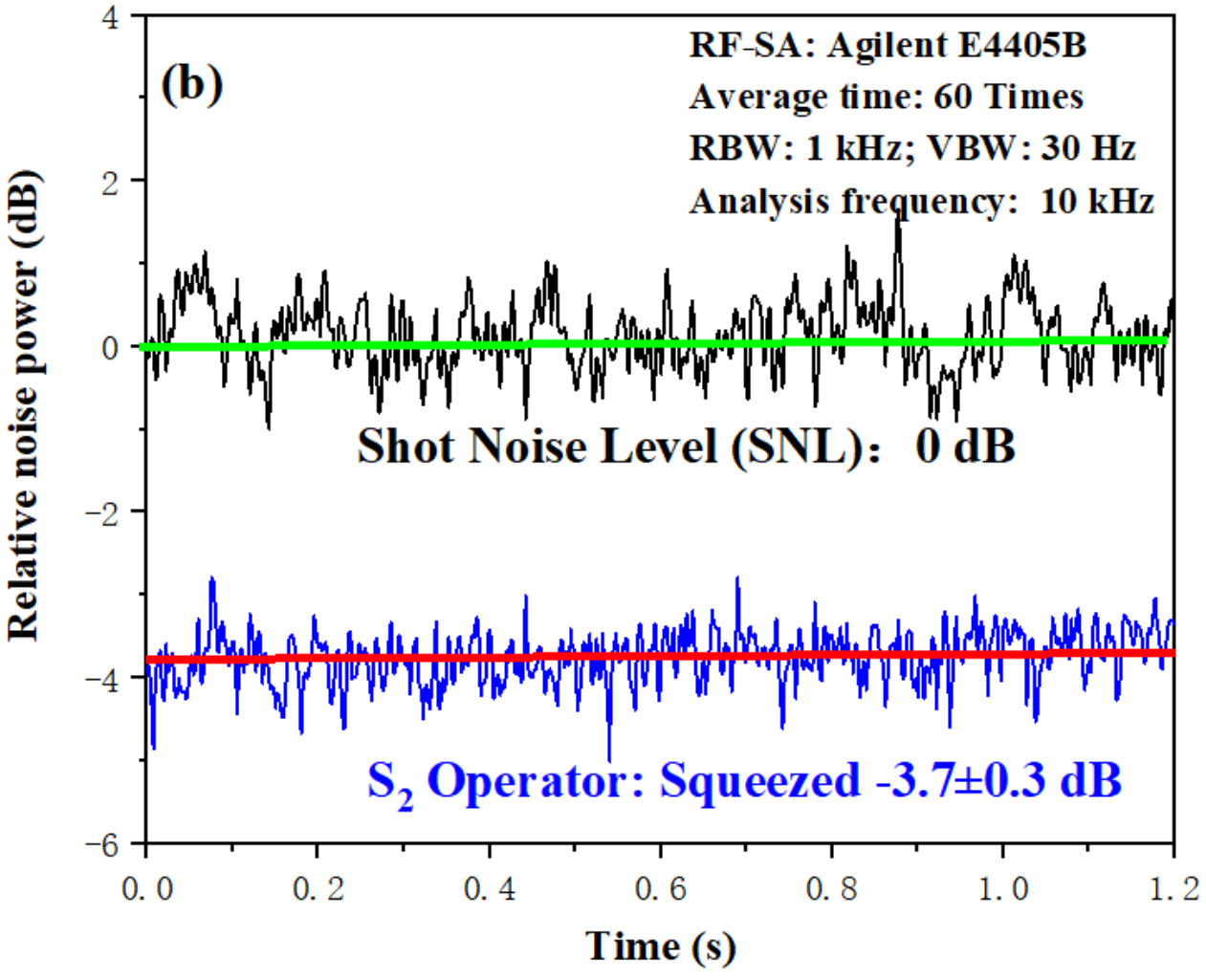}}
\vfill
\subfigure[]{\label{fig:subfig:a}
\includegraphics[scale=0.42]{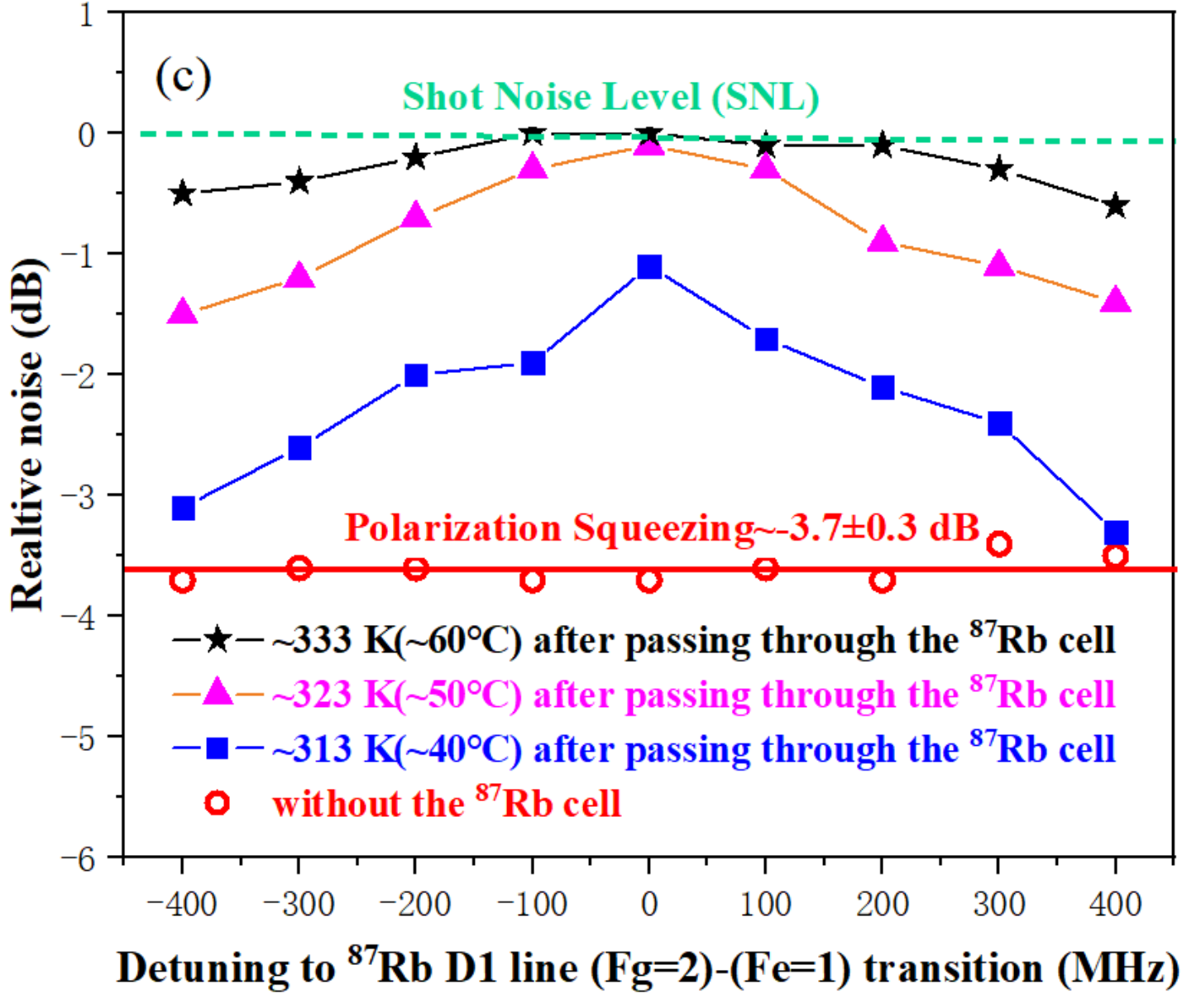}}
\hspace{0.01\linewidth}
   \caption{Models of two light fields (PCS and PSS) on poincare spheres and their measurement results of SNS. (a) SNS signals are measured by PCS and PSS, respectively. The abscissa and ordinate represent the analysis frequency and noise power respectively. (b) Red sphere represents PCS state, and its polarization will rotate around the equator and has a rotation angle change of $\varphi _{FR(PCS)}$ in the measurement of SNS. The blue ellipsoid represents PSS state, whose Stokes operator$\hat{S}{}_{2}$is squeezed and Stokes operator$\hat{S}{}_{3}$is anti-squeezed. It produces a rotation angle change of $\varphi _{FR(PSS) }$under the same condition.}
\end{figure}

As the absorption of laser at near-resonant frequency and the back-action from atoms may dramatically destroy the quantum correlations between photons[26-28], which would be influenced by the laser frequency detuning and atomic number density. As shown in figure 3 (c), the laser frequency is tuned to within the range of $\pm$400 MHz relative to the 5S$_{1/2}$ (F$_g$=2)- 5P$_{1/2}$ (F$_e$=1) transition of D1 line of $^{87}$Rb, and squeezing levels of PSS after passing through the atomic vapor cell with different atomic number density are measured. The red circles represent the squeezing level of about -3.7$\pm$0.3 dB before passing through the atomic ensemble. The blue squares are the squeezing levels after passing through the atomic vapor cell with atomic number density of $\sim$ 5.8×10$^{10}$/cm$^3$ (T$\sim$40 ℃). Obviously, the squeezing level is reduced by 0.5 dB only when the detuning frequency is $\sim$ $\pm$400 MHz. With the decrease of laser frequency detuning, the destruction of quantum properties become more intense, resulting in a significant decrease of squeezing level, especially when resonant frequency is reached, the entanglement and correlation between the twin photons are damaged severely, and corresponding squeezing level is only -1.2$\pm$0.3 dB. However, the pink triangles and the black pentagonal stars represent the squeezing levels after passing through the atomic vapor cell with atomic number density of $\sim$1.5×10$^{11}$/cm$^3$ (T$\sim$50 ℃) and $\sim$3.4×10$^{11}$/cm$^3$ (T$\sim$60 ℃), respectively, which show that corresponding squeezing level is more severely damaged, especially at the resonant frequency.

The core of this work is to improve the sensitivity of magnetic fields by using of 795-nm Stokes operator squeezed light. For our single-beam atomic magnetometer based on Faraday rotation, the atomic number density is controlled at $\sim$ 5.8×10$^{10}$/cm$^3$ (T$\sim$ 40 ℃), which can ensure that the quantum properties of the squeezed light with different laser detunings will not be destroyed completely by atomic ensemble, so as to highlight the quantum enhanced effect in different degrees. 

If we continue to increase the output of the squeezing level to a certain extent, the sensitivity will be more higher. Predictably, according to the result of squeezing level at 860 nm [29], a higher squeezing level of -9 dB at 795 nm, even up to -10 dB, is expected to be achieved by further suppressing the phase noise of the laser, reducing the cavity loss in the OPO pr℃ess, and searching for crystals with higher nonlinear coefficient, lower loss and weak absorption effect. In our experiment, assuming that the loss of the light field due to the absorption of atoms is constant, the reaction of the spin noise of the atomic ensemble coupled to the $\widehat{S}_2$ component of the PSS is also certain. Therefore, for the input PSS of light field with different squeezing levels, the background noise of the atomic magnetometer will be improved to different degrees after the interaction between the atomic ensemble and the optical field. In this work, the background noise of the atomic magnetometer is improved by 3.2 dB when the PSS of light with the squeezing level of -3.7 dB and the laser detuning of -400 MHz relative to the 5S$_{1/2}$ (F$_g$=2) - 5P$_{1/2}$ (F$_e$=1) transition of  $^{87}$Rb D$_1$ line is input into our atomic magnetometer. That is to say, the PSS of light with squeezing level of -3.2 dB can be measured after passing through the atomic magnetometer. However, if we assume that the squeezing level of the input PSS of field is $\sim$ -10 dB, under the same conditions, the squeezing level after passing through the atomic magnetometer will be reduced to $\sim$ -7.4 dB [30]. In other words, the greater the squeezing level produced by the OPO, the more severe the influence or destruction from the atomic ensemble on the squeezing level, so that the measured squeezing will eventually converge to a level. Thus, increasing the produced squeezing indefinitely, or applying greater squeezing, will not lead to greater measured squeezing improvement. In order to fully use the PSS’ quantum characters in optically pumped atomic magnetometer, we must firstly reduce the absorption losses and backaction from atomic spin noise to the Stokes operater $\widehat{S}_2$ polarization squeezing.

\section{Quantum-enhanced $^{87}$Rb magnetometer based on Faraday rotation and discussions}
The response of atomic magnetometer to external magnetic field depends on the strength of light-atom interaction, which is related to laser intensity, laser frequency detuning, and atomic number density. We measured the relationship between the rotation angle $\varphi$ of the polarization plane and the magnetic field (-6000 pT $\sim$ +6000 pT) with the laser intensity of 3.5 mW/cm$^2$ when the laser frequency is locked to the (F$_g$=1) - (F$_e$=1), (F$_g$=1) - (F$_e$=2), (F$_g$=2) - (F$_e$=1), and (F$_g$=2) - (F$_e$=2) hyperfine transitions of D$_1$ line (795 nm), respectively. Among them, the greater the absolute value of the slope near the zero magnetic field, the more sensitive it is to the change of external magnetic field.
\begin{figure}[htbp!]
\centering
\subfigure[]{\label{fig:subfig:a}
\includegraphics[scale=0.19]{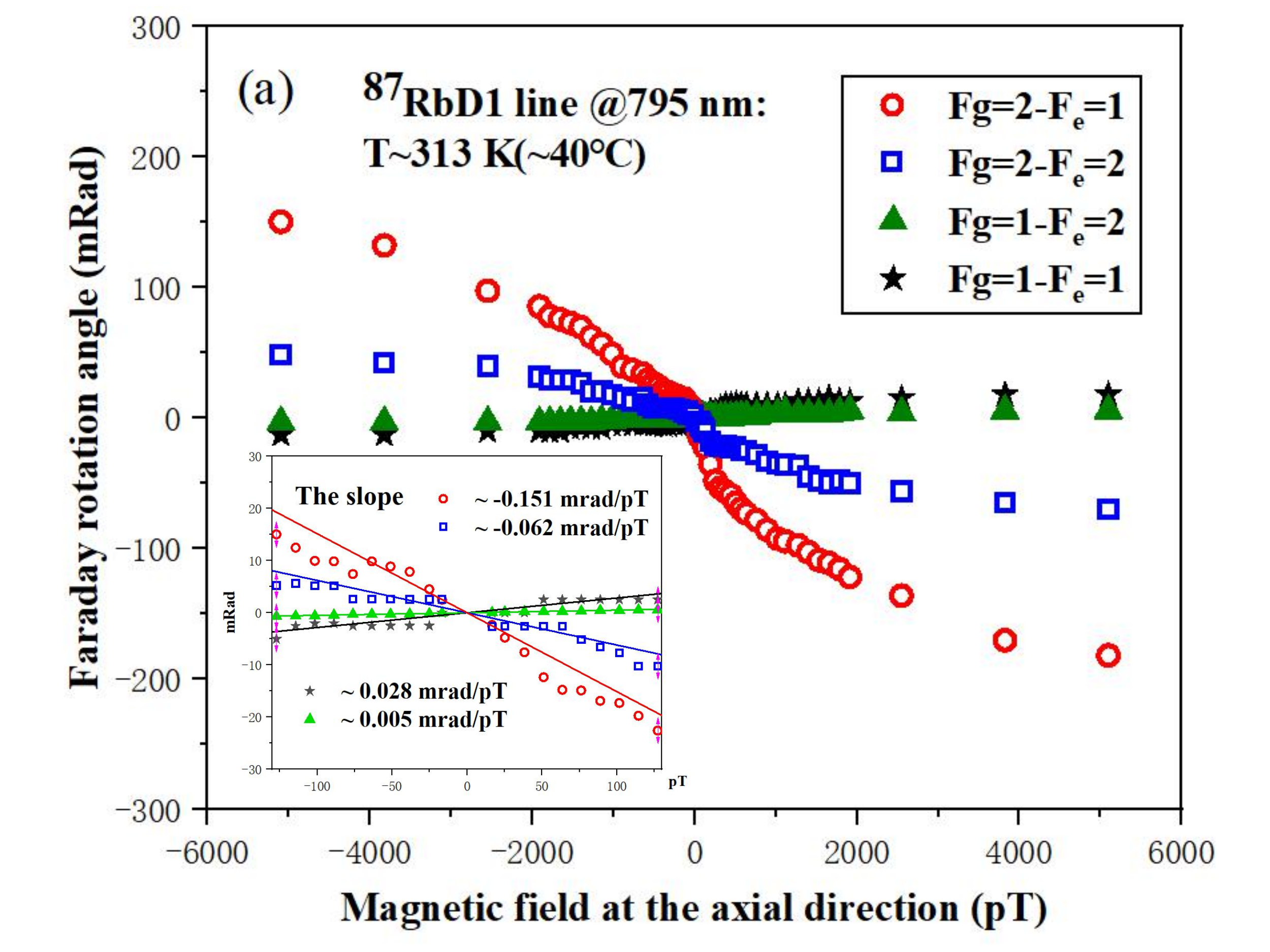}}
\hspace{0.01\linewidth}
\subfigure[]{\label{fig:subfig:b}
\includegraphics[scale=0.19]{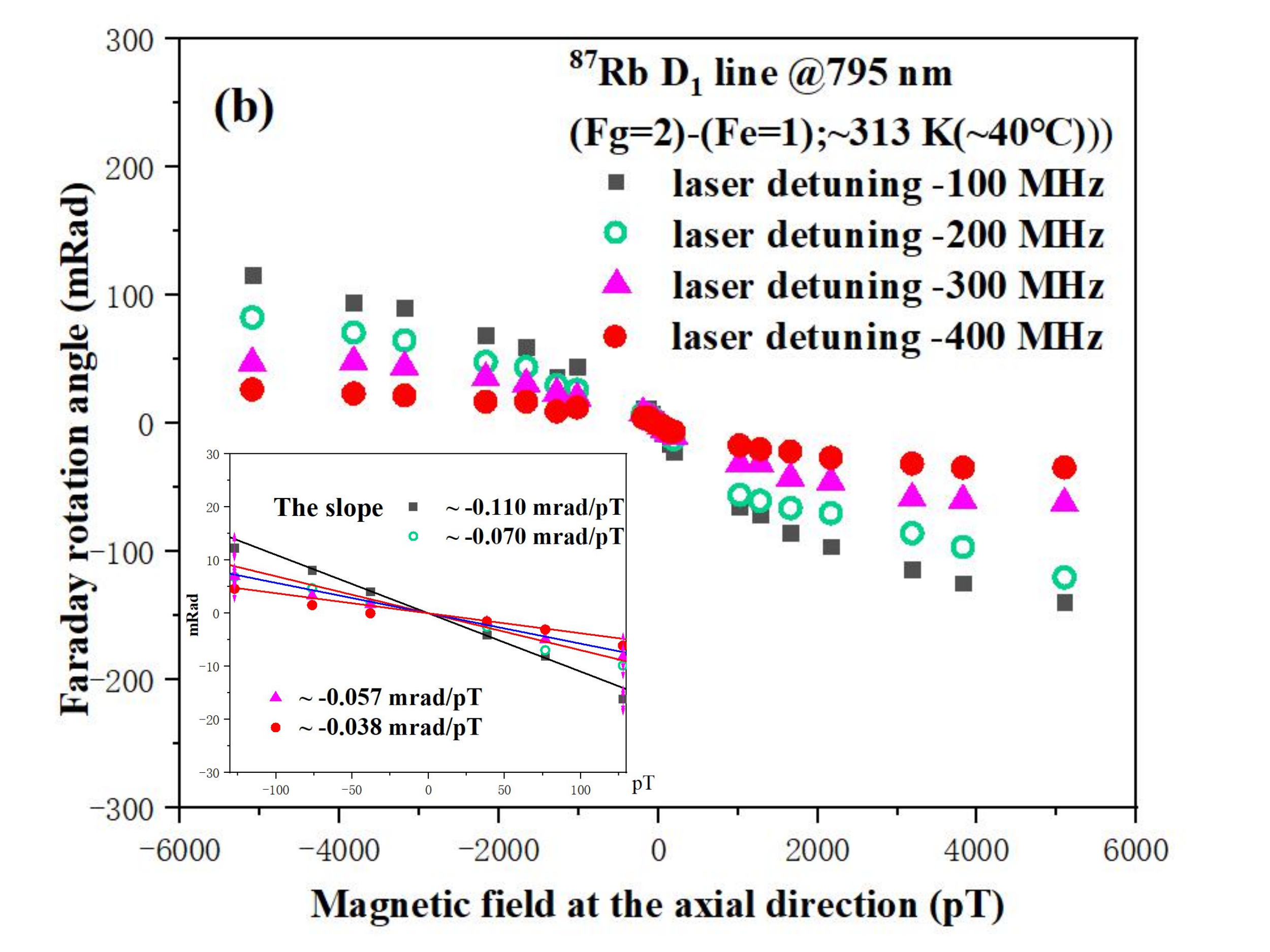}}
\vfill
\subfigure[]{\label{fig:subfig:a}
\includegraphics[scale=0.19]{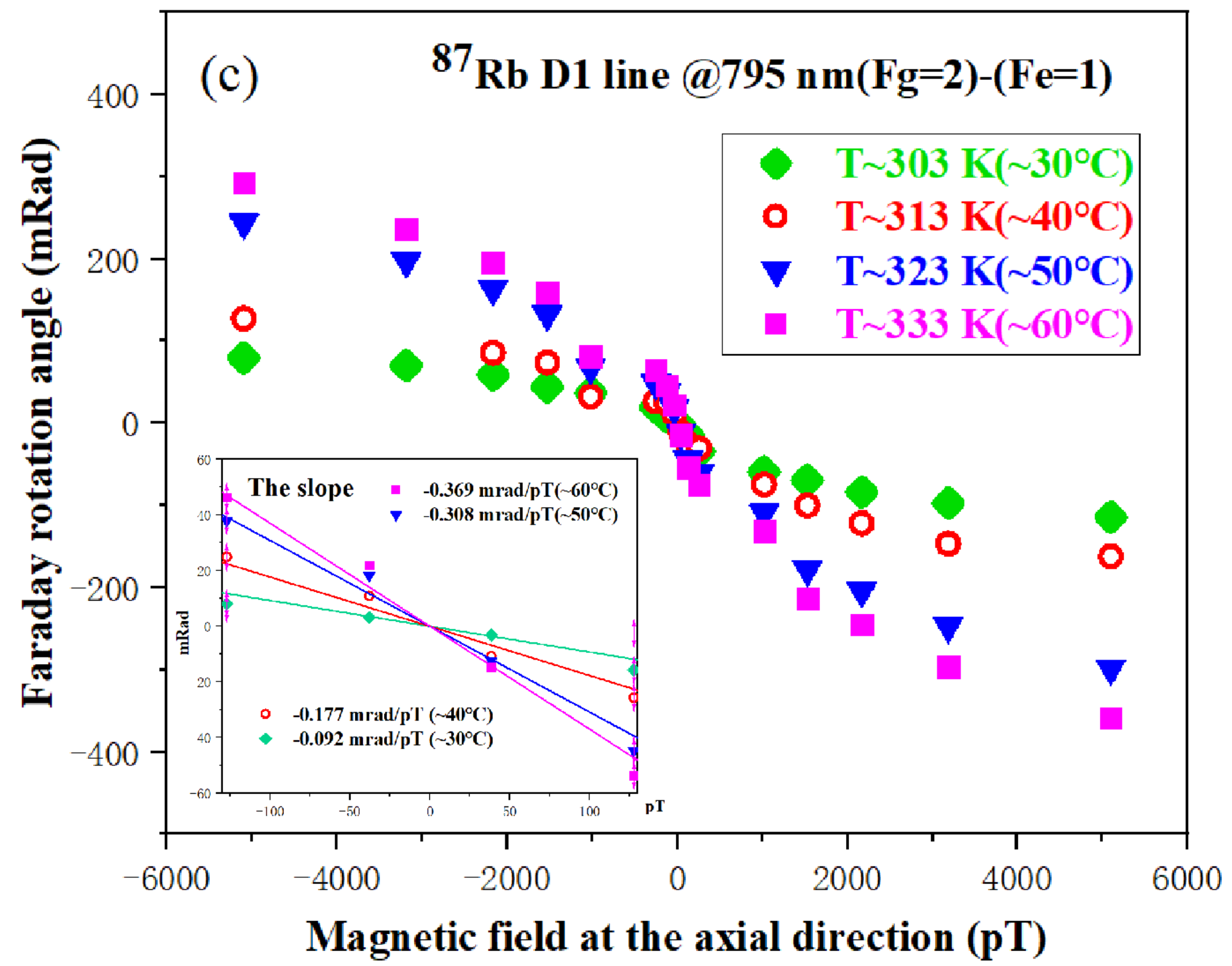}}
\hspace{0.01\linewidth}
   \caption{(a) Relationship between the polarization rotation angles ($\varphi$) and the applied axial magnetic fields with different laser’s frequency. The red circles, blue empty squares, green triangles, and black stars represent the cases for the laser frequency locked to the (F$_g$=2) - (F$_e$=1), (F$_g$=2) - (F$_e$=2), (F$_g$=1) - (F$_e$=2), and (F$_g$=1) - (F$_e$=1) transitions of $^{87}$Rb D$_1$ line (795 nm), respectively. The inset represents the sensitivity of Faraday rotation angle under each laser’s frequency and the solid lines are corresponding linear fits, which are -0.151 mrad/pT, -0.062 mrad/pT, 0.005 mrad/pT, 0.028 mrad/pT, respectively. (b) Dependence of Faraday rotation angle on laser frequency detuning relative to the 5S$_{1/2}$ (F$_g$=2) - 5P$_{1/2}$ (F$_e$=1) transition of D1 line of $^{87}$Rb. The black squares, the green hollow circles, the pink triangles and the red solid circles represent the laser frequency detuning of -100 MHz, -200 MHz, -300 MHz, -400 MHz, respectively. The inset represents the sensitivity of Faraday rotation angle under each laser frequency detuning and the solid lines are corresponding linear fits, which are -0.110 mrad/pT, -0.070 mrad/pT, -0.057 mrad/pT, -0.038 mrad/pT, respectively. (c) Influence of atomic number density on Faraday rotation angle, which the green diamonds, red circles, blue triangles, and pink squares represent the cell’s temperature of 303 K ($\sim$ 30℃), 313 K ($\sim$ 40℃), 323 K ($\sim$ 50℃), 333 K ($\sim$ 60℃), respectively. The inset represents the sensitivity of Faraday rotation angle under atomic vapor cell with different temperatures and the solid lines are corresponding linear fits, which are -0.092 mrad/pT, -0.205 mrad/pT, -0.329 mrad/pT, -0.398 mrad/pT, respectively.}
\end{figure}

Compared with the hyperfine ground state (F$_g$=1), Faraday rotation angle corresponding to hyperfine ground state (F$_g$=2) is significantly increased, and the slopes are opposite, which is mainly attributed to the different refractive index of atomic media and the g factor in the hyperfine structures. The laser frequency resonant with the 5S$_{1/2}$ (F$_g$=2) - 5P$_{1/2}$ (F$_e$=1) transition of D$_1$ line of $^{87}$Rb can be the best choice as the maximal slope with the absolute value of $\sim$ 0.157 mrad/pT, as shown by the red circles in figure 4 (a). With the increase of laser frequency detuning, the spin polarization of atoms induced by the laser is weakened, the slope of magnetically induced Faraday rotation angle decreases, as shown in figure 4 (b). In addition, we also verified that the response sensitivity of the system to the external magnetic field is positively correlated with the atomic number density within a certain range, as shown in figure 4 (c). Taking all these into account, for the quantum-enhanced rubidium atomic magnetometer system, the advantages of near resonant laser and high atomic number density cannot be exploited simultaneously, and in order to demonstrate the characteristics of PSS at high atomic number density, we set the vapor cell’s temperature at $\sim$ 313 K (T$\sim$ 40 ℃).
\begin{figure}
\centering
\subfigure[]{\label{fig:subfig:a}
\includegraphics[scale=0.4]{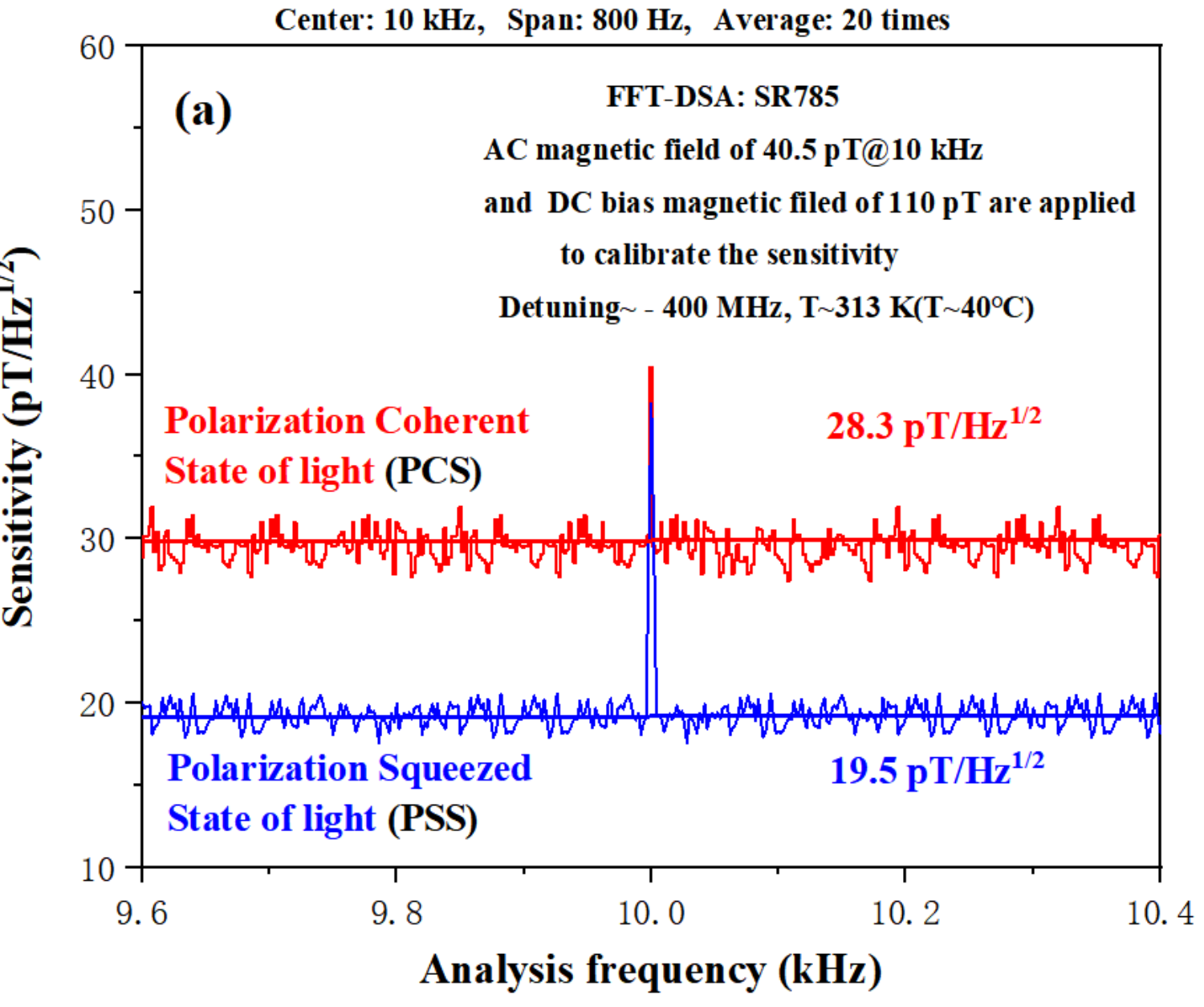}}
\hspace{0.01\linewidth}
\subfigure[]{\label{fig:subfig:b}
\includegraphics[scale=0.4]{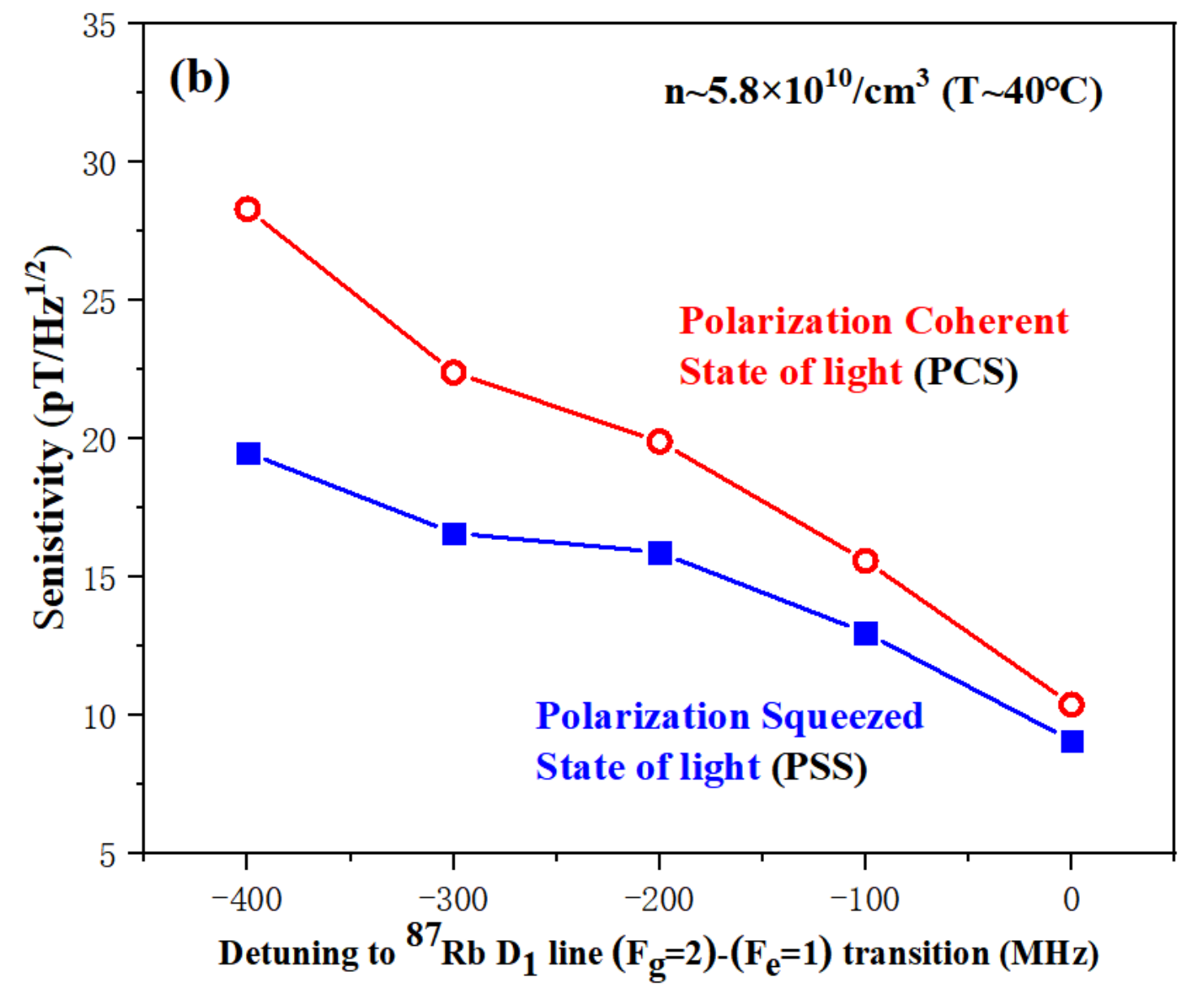}}
\vfill
   \caption{(a) Normalized noise power spectra measured by FFT dynamic signal analyzer (Model SR-785, SRS Inc). An AC magnetic field of 40.5 pT at 10 kHz and a DC bias magnetic field of 110 pT are applied to calibrate the sensitivity. The red line represents the results of PCS, while the blue line represents the results of PSS. Employing -3.7$\pm$0.3 dB of Stokes operator $\widehat{S}_2$ squeezed light with a frequency detuning of $\sim$ -400 MHz relative to the 5S$_{1/2}$ (F$_g$=2) - 5P$_{1/2}$ (F$_e$=1) transition of D$_1$ line of $^{87}$Rb, the sensitivity of magnetic field measurement is improved from 28.3 pT/Hz$^{1/2}$ to 19.5 pT/Hz$^{1/2}$. (b) Sensitivity of magnetic field measurement in the cases of PSS (blue triangles) and PCS (red circles) with different laser detuning. The atomic number density is $\sim$ 5.8×10$^{10}$/cm$^3$ (T$\sim$ 40 ℃), and the zero frequency detuning means that light is resonant to the 5S$_{1/2}$ (Fg=2) - 5P$_{1/2}$ (F$_e$=1) transition of $^{87}$Rb D$_1$ line.}
\end{figure}

Under a condition of DC bias magnetic fields of 110 pT, the AC calibrated field of 40.5 pT at 10 kHz, and the atomic number density of $\sim$ 5.8×10$^{10}$/cm$^3$ (T$\sim$40 ℃), for a probe light with laser frequency detuning of $\sim$ -400 MHz relative to the 5S$_{1/2}$ (Fg=2) - 5P$_{1/2}$ (F$_e$=1) transition of $^{87}$Rb D$_1$ line, the sensitivity has been preliminarily enhanced from $\sim$28.3 pT/Hz$^{1/2}$ to $\sim$19.5 pT/Hz$^{1/2}$, as shown in figure 5(a). We also analyzed the relationship between the sensitivity of magnetic field measurement and the laser frequency detuning under the probe of PSS and PCS, shown in figure 5(b). The red circles and blue squares connected by solid lines represent the sensitivity of magnetic fields under PCS and PSS, respectively. With the increase of laser frequency detuning, the sensitivity gets worse comparing with the zero-detuning case, but it can fully embody the PSS’s quantum enhanced properties due to the maintained PSS with high squeezing level. On the contrary, although the absolute sensitivity under the action of PCS is the highest when the laser frequency resonates with the atomic transition (zero-detuning case), it is difficult to reflect the quantum enhanced effect with reserved squeezing level of only -1.2 dB due to the serious deterioration of the quantum properties. We believe that when the laser frequency detuning is arbitrary under the condition of atomic number density less than $\sim$ 5.8×10$^{10}$/cm$^3$ (T$\sim$ 40 ℃) and within a limited range of squeezing levels, the PSS with higher squeezing level is more conducive to the enhancement of sensitivity, even though the destruction of PSS’s quantum characteristic after passing though the atomic magnetometer increases with the imput squeezing level. It's worth noting that if the atomic number density is increased to $\sim$ 2.2×10$^{11}$/cm$^3$ (T$\sim$ 55℃), the damage of atoms to the light field will be more severe, and as a result, the noise background under the measurement of PSS will be higher than that of the PCS, and corresponding sensitivity will be reduced, which has been proved in reference [4].

Although the PSS’s quantum properties are somewhat weaken due to the absorption loss under different laser frequency detuning, our $^{87}$Rb atomic magnetometer with PSS as probe always shows quantum enhancement at the same conditions, as shown in Fig. 5(a), comparing the case with PCS as probe. By further optimizing the technology and scheme to reduce the damage of quantum properties for example, one laser is used for pumping and another large detuned laser is used for detection.

\section{Conclusions}
In this work, we analyzed the relationship between laser frequency detuning, squeezing level and magnetic field sensitivity. The results show that the larger laser frequency detuning is, the higher the maintained squeezing level of PSS is, and the more obvious the quantum enhancement effect is; on the contrary, the smaller the laser frequency detuning is, the higher the absolute sensitivity of magnetic field is, and the weaker the quantum enhancement effect is. 

For our FR-type magnetometer, in order to improve the sensitivity of magnetic field measurement, the probe laser has to be tuned to the near resonant laser frequency range at the expense of the destruction of the PSS. In this case, the atomic ensemble can generate magnetic quadrupole moment to some extent, in another word, the realization of atomic alingment, which will lead to a higher slope of rotation angle of the polarization plane caused by magnetically induced Faraday rotation, and its contribution to the sensitivity is greater than the negative impact due to the absorption (scattering) losses of the light field under the resonant or near-resonant conditions. This maybe one of   main reasons for our sensitivity of several tens pT/Hz$^{1/2}$. However, when PSS of light resonates or near-resonates with atomic transition, the absorption (scattering) losses by atoms not only directly reduce the intensity of transmitted light field, but also partially damage the quantum characters of PSS. Especially, the spin noise of atoms can couple with the PSS through absorption process, which has backaction on the PSS, therefor, the quantum characters of PSS are destroyed to a large extent. In this case, the sensitivity with PSS is almost not different compared to the sensitivity with PCS, and the quantum enhancement effect is limited. However, if the squeezing level of PSS is higher within a certain range and the loss of the system is further reduced, it is entirely possible that the magnetic field sensitivity measured by PSS at far detuning frequency is even better than that measured by PCS at the resonant frequency.

From the best perspective, in order to make full use of the advantages of PSS and high atomic number density, an another resonant circularly polarized beam can be used as pump laser to polarize the atomic ensemble in the next scheme [31], while PSS state of light with large frequency detuning in the direction perpendicular to the pump laser is introduced to be a probe laser, so the damage to the PSS’ quantum characteristics caused by the atomic ensemble can be significantly reduced.

\textbf{Acknowledgments:}
This work is financially supported by the National Natural Science Foundation of China (Grant Nos. 11974226, 61905133, 11774210, and 61875111), the National Key R\&D Program of China (Grant No. 2017YFA0304502), Shanxi Provincial Graduate Education Innovation Project (Grant No. 2020BY024) and the Shanxi Provincial 1331 Project for Key Subjects Construction. 
 
\textbf{Conflict of interest:} The authors declare that they have no conflict of interest.


\begin{thebibliography}{}
\bibitem{1} Tse M, et al., Quantum-enhanced advanced LIGO detectors in the era of gravitational-wave astronomy, Phys. Rev. Lett. 123 (23), 2331107 (2019)
\bibitem{2} Acernese F, et al., Increasing the astrophysical reach of the advanced VIRGO detector via the application of squeezed vacuum states of light, Phys. Rev. Lett. 123 (23), 2331108 (2019)
\bibitem{3} Wolfgramm F, Cerè A, Beduini A B, Predojević A, Koschrreck M, and Mitchell M, Squeezed-light optical magnetometry, Phy. Rev. Lett. 105 (5), 053601 (2010)
\bibitem{4} Horrom T, Singh R, Dowling J P, and Mikhailov E E, Quantum-enhanced magnetometer with low-frequency squeezing, Phys. Rev. A 86 (2), 023803 (2012)
\bibitem{5} Novilova I, Mikhailov E E, and Xiao Y H, Excess optical quantum noise in atomic sensors, Phys. Rev. A 91 (5), 051804 (2015)
\bibitem{6} Otterstrom N, Pooser R C, and Lawrie B J, Nonlinear optical magnetometry with accessible in situ optical squeezing, Opt. Lett. 39 (22), 6533 (2014)
\bibitem{7} Li B B, Bilek J, Hoff U B, Madsen L S, Forstner S, Prakash V, Schäfermeier C, Gehring T, Bowen W P, and Andersen U L, Quantum enhanced optomechanical magnetometry, Optica 5 (7), 850 (2018)
\bibitem{8} Lucivero V G, Jiménez-Martínez R, Kong J, and Mitchell M, Squeezed-light spin noise spectroscopy, Phys. Rev. A 93 (5), 053802 (2016)
\bibitem{9} Vasilakis G, Shen H, Jensen K, Balabas M V, Salart D, Chen B, and Polzik E S, Generation of a squeezed state of an oscillator by stroboscopic back-action-evading measurement, Nature Phys.11 (5), 389 (2015)
\bibitem{10} Bao H, Duan J L, Jin S C, Lu X D1, Li P X, W Z,,Wang M F, Novikova I, Mikhailov E E, Zhao K F, Mølmer K, Shen H, and Xiao Y H, Spin squeezing of 1011 atoms by prediction and retrodiction measurements, Nature 581 (5), 159 (2020)
\bibitem{11} Dang H B, Maloof A C, and Romalis M V, Ultrahigh sensitivity magnetic field and magnetization measurements with an atomic magnetometer, Appl. Phys. Lett. 97 (15), 151110 (2010)
\bibitem{12} Stahler M, Knappe S, Affolderbach C, Kemp W, and Wynands R, Picotesla magnetometry with coherent dark states. Europhys. Lett. 54 (3), 323 (2001)
\bibitem{13} Mckenzie K, Mikhailov E E, Goda K, and Ping K L, and Mcclelland D E, Quantum noise locking. J. Opt. B: Quant. \& Semiclass. Opt. 7(10). S421 (2005)
\bibitem{14} Bowen W P, Schnabel R, Bachor H A, and Ping K L, Polarization squeezing of continuous variable Stokes parameters, Phys. Rev. Lett. 88 (9), 093601 (2002)
\bibitem{15} Schnabel R, Bowen W P, Treps N, Ralph T C, Bachor H A, and Lam P K, Stokes-operator-squeezed continuous-variable polarization states, Phys. Rev. A 67 (1), 012316 (2003)
\bibitem{16} Qin Z Z, Gao L M, Wang H L, Marino A M, Zhang W P, and Jing J T, Experimental generation of multiple quantum correlated beams from hot rubidium vapor, Phys. Rev. Lett. 113 (2), 023602 (2014)
\bibitem{17} Fang Y M and Jing J T, Quantum squeezing entanglement from a two-mode phase-sensitive amplifier via four-wave mixing in rubidium vapor, New J. Phys. 17 (2), 023027 (2015)
\bibitem{18} Han Y S, Wen X, He J, Yang B D, Wang Y H, and Wang J M, Improvement of vacuum squeezing resonant on the rubidium D1 line at 795 nm, Opt. Express 24 (3), 2350 (2016)
\bibitem{19} Wen X, Han Y S, Liu J Y, He J, and Wang J M, Polarization squeezing at the audio frequency band for the rubidium D1 line, Opt. Express 25 (17), 20737 (2017)
\bibitem{20} Gerginov V, Da Silva F C S, and Howe D, Prospects for magnetic field communications and l℃ation using quantum sensors, Rev. Sci. Instr. 88 (12), 125005 (2017)
\bibitem{21} Lucivero V G, Anielski P, Gawlik W, and Mitchell M, Shot-noise-limited magnetometer with sub-picotesla sensitivity at room temperature. Rev. Sci. Instr. 85 (11), 113108 (2014)
\bibitem{22} Allred J C, Lyman R N, Kornack T W, and Romalis M V, High-sensitivity atomic magnetometer unaffected by spin-exchange relaxation, Phys. Rev. Lett. 89 (13), 130801 (2002)
\bibitem{23} Kominis I K, Komack T W, Allred J C, and Romalis M V, A sub-femtotesla multichannel atomic magnetometer, Nature 422 (6932), 596 (2003)
\bibitem{24} Wen X, Han Y S, Liu J Y, Bai L L, He J, and Wang J M, Generation of squeezed states at low analysis frequencies, Acta Physica Sinica 67 (2), 024207 (2018) (in Chinese)
\bibitem{25} Bai L L, Wen X, Yang Y L, He J, and Wang J M, Laser intensity noise suppression for preparing audio-frequency squeezed vacuum state of light, Appl. Sci. 10 (4), 1415 (2020)
\bibitem{26} Auzinsh M, Budker D, Kimball D F, R℃hester S M, Stalnaker J E, Sushkov A O, and Yashchuk V V, Can a quantum nondemolition measurement improve the sensitivity of an atomic magnetometer? Phys. Rev. Lett. 93 (17), 173002 (2004)
\bibitem{27} Wasilewski W, Jensen K, Krauter H, Renema J J, Balabas M V, and Polzik E S, Quantum noise limited and entanglement-assisted magnetometry, Phys. Rev. Lett. 104 (13), 133601 (2010)
\bibitem{28} Koschorreck M, Napolitano M, Dubost B, and Mitchell M, Sub-projection-noise sensitivity in broadband atomic magnetometry, Phys. Rev. Lett. 104 (9), 093602 (2010)
\bibitem{29} Takeno Y; Yukawa M; Yonezawa H; Furusawa A, Observation of -9 dB quadrature squeezing with improvement of phase stability in homodyne measurement, Opt. Express 15 (7), 4321 (2007)
\bibitem{30} Chua S S Y, Slagmolen B J J, Shadd℃k D A and McClelland D E, Class. Quantum Grav., 31 (18), 183001 (2014)
\bibitem{31} Deans C, Marmugi L, and Renzoni F, Sub-picotesla widely tunable atomic magnetometer operating at room-temperature in unshielded environments, Rev. Sci. Instr., 89 (8), 083111 (2018)


\end{thebibliography}
\end{document}